\documentstyle[12pt,bezier]{article}

\def\ltap{\raisebox{-.55ex}{\rlap{$\sim$}} \raisebox{.4ex}{$<$}}

\def\lsim{\mathrel{\ltap}}

\def\e{\mbox{e}}

\begin{document}
\title{Grand unification and heavy axion}
\author{
   V.~A.~Rubakov\\
    {\small \em Institute for Nuclear Research of the Russian Academy of
  Sciences,}\\
  {\small \em 60th October Anniversary prospect 7a, Moscow 117312}\\
  }
\maketitle
\begin{abstract}
We argue that sufficiently complex grand unified theories involving
extra strong intractions that confine at very short distances, may lead 
to a heavy axion solution of the CP problem of QCD. This axion
may have
a mass within accessible energy range, and its low energy interactions 
emerge through mixing with axial Higgs boson(s). Another signature of 
this scenario is softly broken Peccei---Quinn symmetry in the electroweak
Higgs sector. We present a toy GUT exhibiting these features.

\end{abstract}

\vskip .5 in



In QCD, the effective $\theta$-parameter 
$\bar{\theta} = \theta + \mbox{arg~det} M_{quark}$ breaks CP
\cite{Hooft} and is experimentally constrained to be unnaturally
small, $\bar{\theta} \lsim 10^{-10}$ (for reviews see, e.g., 
ref. \cite{Reviews}). An elegant solution to this strong CP problem
is based on the Peccei---Quinn (PQ) symmetry \cite{PQ} and predicts a light 
particle, axion. In view of constraints from experimental searches
for the Weinberg---Wilczek \cite{WW} axion, a widely accepted option 
is an extremely light invisible axion \cite{invisible}. A potential 
problem with the latter comes from possible non-renormalizable terms 
in the low energy Lagrangian which may be due to very high (say, Planckian)
scales and need not respect PQ symmetry \cite{gravity}. Negligible for other
purposes, these terms would introduce extra axion potential and ruin the
Peccei---Quinn mechanism precisely because the QCD contribution into
the axion potential is tiny. From this point of view it is more 
safe to have axion heavy enough.

In this paper we point out that heavy axions may appear in sufficiently
complex grand unified theories containing extra gauge interactions
(with unbroken gauge group) which become strong well above the
accessible energies. The effective $\theta$-parameters of QCD and these 
extra strong interactions may be equal to each other due to a symmtry
built in a GUT. If there is {\it one} PQ symmetry relevant to {\it both} 
QCD and extra strong interactions, the Peccei---Quinn mechanism rotates away
both of these $\theta$-parameters, while axion obtains its mass predominantly
from extra strong interactions and is therefore heavy (much heavier
than  Weinberg---Wilczek one).

Similar idea was put forward by Tye \cite{Tye} 
(see also ref.\cite{Krasnikov})
in the context of technicolor
plus Higgs models with PQ symmetry. However, many (if not all) such models
predict numerous pseudo-Goldstone bosons; some of these are charged and have
masses well below 40 GeV, which is ruled out experimentally. This problem 
is not inherent in GUTs\footnote{Mechanisms that may make invisible
axion heavy enough are discussed in ref. \cite{Peskin}}.

To be specific, let us consider a toy GUT. This model will not be 
realistic for several reasons, but we expect that it picks up generic
features of possible heavy axions. The model is a non-supersymmetric GUT
with the gauge group $SU(5)\times SU(5)$, where the first $SU(5)$ is 
meant to model real world and the second $SU(5)$ is a mirror group. Let the
fermionic and Higgs content be mirror symmetric. Ordinary (mirror) fermions
are singlets under mirror (ordinary) $SU(5)$ and form the usual $\bar{5}$- 
and 10-plets under ordinary (mirror) $SU(5)$. There is one Higgs 
24-plet and two Higgs 5-plets in each $SU(5)$ sector which are singlets 
under partner $SU(5)$. Let us require that at this stage each $SU(5)$ sector 
has its own PQ symmetry that rotates the two Higgs 5-plets in the 
opposite ways,
\begin{equation}
 \varphi_5^{(1)} \to \e^{i\alpha} \varphi_5^{(1)}, ~~~
\varphi_5^{(2)} \to \e^{-i\alpha} \varphi_5^{(2)}
\label{4*}
\end{equation}
for ordinary Higgs 5-plets $\varphi_5^{(1,2)}$, and
\begin{equation}
 \Phi_5^{(1)} \to \e^{i\beta} \Phi_5^{(1)}, ~~~
\Phi_5^{(2)} \to \e^{-i\beta} \Phi_5^{(2)}
\label{4**}
\end{equation}
for mirror Higgs 5-plets $\Phi_5^{(1,2)}$. 
To have just one PQ symmetry, let us introduce 
an $SU(5)\times SU(5)$ singlet complex scalar field $S$ of PQ charge 1
that interacts with both ordinary and mirror Higgs 5-plets,
\begin{equation}
 L_{S,\varphi, \Phi} = h \varphi_5^{(1) \dagger} \varphi_5^{(2)}S^2 +
                       h' \Phi_5^{(1) \dagger} \Phi_5^{(2)} S^2 
                        + \mbox{h.c.}
\label{4+}
\end{equation}
The self-interaction of $S$ is required to be symmetric under the 
phase rotations of $S$, so the remaining PQ symmetry is (\ref{4*}), 
(\ref{4**}) with $\beta = \alpha$ and $S \to \e^{i\alpha} S$. Let us 
assume for definiteness that $S$ does not obtain vacuum expectation value,
though this assumption is not crucial for further discussion.

Let us now require that at the (high) energy scales where
both $SU(5)$ groups are unbroken,
hard (dimension 4) terms in the whole Lagrangian 
are mirror symmetric, while the soft terms are not. This implies, in
particular, that $\theta_{ordinary} = \theta_{mirror}$
(the $\theta$-terms are hard) and that the phases of Yukawa couplings are 
the same in ordinary and mirror sectors. This requirement also implies the
equality of the couplings entering eq.(\ref{4+}), $h'=h$. Hence, without 
loss of generality one sets $h$ to be real (the phase of $h$ can be rotated 
away by the phase rotation of $S$). In one loop, the interaction (\ref{4+})
introduces direct interaction between ordinary and mirror Higgs 5-plets,
\begin{equation}
  L_{\varphi, \Phi} = \lambda \left( \varphi_5^{(1) \dagger} \varphi_5^{(2)}
                               \right) \cdot
                       \left( \Phi_5^{(2) \dagger} \Phi_5^{(1)}
                               \right) + \mbox{h.c.}
\label{5+}
\end{equation}
where $\lambda \propto h^2 \ln (m_S/\mu)$ and $\mu$ is the normalization 
scale. Note that $\lambda$ is real and the interaction (\ref{5+}) is still
PQ symmetric.

   Let us require that, just like ordinary  $SU(5)$, mirror $SU(5)$ breaks
down to $SU(3)_{mc}\times U(1)_{mEM}$, where mc and mEM refer
to mirror color and mirror electromagntism, respectively. Since the soft
terms of the mirror sector are different from those of ordinary sector, 
this breaking occurs at different energy scales. Consider the case when 
mirror $SU(5)$ breaks down at much lower energy than the ordinary GUT scale. 
The coupling constant of $SU(5)$ runs faster than that of $SU(3)$, so
the mirror coupling constant is larger than that of ordinary $SU(3)_c$
at the point where mirror $SU(5)$ breaks down. Hence, $SU(3)_{mc}$
becomes strong at the scale $\Lambda_{mc}$ which is larger than
ordinary $\Lambda_{QCD}$. Assuming
\begin{equation}
   <\Phi_5^{(1)}> \sim <\Phi_5^{(2)}> \sim v_m > \Lambda_{mc}
\label{6*}
\end{equation}
we have the mirror world similar to the ordinary world, but scaled up in 
energy (and with $v_m/\Lambda_{mc}$ not necessarily of the same
order as the ratio of the ordinary Higgs expectation value to QCD
confinement scale, $v/\Lambda_{QCD} \sim 10^{3}$).

By the mirror symmetry of the hard terms, the effective $\theta$-parameters 
of ordinary and mirror sectors are equal to each other\footnote{Note  
that the soft terms consistent with PQ symmetry
do not contain phases, which otherwise
would be different in ordinary and mirror sectors.}, at least at the tree 
level. By performing
 PQ rotation, one makes both of them equal to zero. By the 
mirror PQ mechanism, the axion field then takes zero vacuum expectation 
value, and both mirror and ordinary strong interactions conserve CP. In
other words, at non-zero mirror effective $\theta$-parameter, 
$\bar{\theta}_{mirror}$, the phase of the vacuum expectation value of 
$\Phi_5^{(1) \dagger} \Phi_5^{(2)}$ is proportional to 
$\bar{\theta}_{mirror}$ by PQ mechanism; the interaction (\ref{5+})
aligns the phase of $\varphi_5^{(1) \dagger} \varphi_5^{(2)}$ to the 
same value, so the effective ordinary $\theta$-parameter, after PQ rotation,
becomes equal to $\bar{\theta}_{ordinary} - \bar{\theta}_{mirror} = 0$, 
at least at the tree level 
\footnote{The effective $\theta$ of ordinary strong interactions may
acquire radiative corrections, but they are small \cite{corrections}.}.

The axion obtains its mass predominantly due to mirror strong interactions.
It is, in fact, a mirror Weinberg---Wilczek axion. The expression for 
the mass is a scaled up version of the Weinberg formula. Recalling that the
mass of the Weinberg---Wilczek axion scales as 
$m_{WW} \propto \Lambda_{QCD}^{3/2} v^{-1/2}$ we estimate the axion 
mass in our model as
\[
  M_a \sim 
  \left(\frac{\Lambda_{mc}}{\Lambda_{QCD}}\right)^{\frac{3}{2}}
   \cdot  \left(\frac{v}{v_{m}}\right)^{\frac{1}{2}}
   \cdot m_{WW}
\]
This may  certainly be much larger that $m_{WW}$.

To get an idea of numbers, let us point out that non-supersymmetric 
$SU(5)$ becomes strong at about $10^5$ GeV. Hence,
$\Lambda_{mc} \lsim 10^5$ GeV. Under the assumption (\ref{6*}) and
using $m_{WW} \sim 100$ KeV, we have
\[
  M_a \lsim 1~\mbox{TeV}
\]
Let us stress that by varying $\Lambda_{mc}$ and $v_{m}$ one can 
easily get the axion much lighter than 1 TeV. Say, at 
$\Lambda_{mc} \sim 3$ TeV and $v_m \sim 10$ TeV one has $M_a \sim 20$ GeV.

The axion interactions with ordinary matter come from the term (\ref{5+}).
At energies below $v_m$ we have
\begin{equation}
 \Phi_5^{(2) \dagger} \Phi_5^{(1)} = c_1 v_m^2 + i c_2 v_m a(x)
\label{8*}
\end{equation}
where $c_1$ and $c_2$ are constants of order 1,  and $a(x)$ is the axion field.
The first term here produces the off-diagonal mass term for the ordinary Higgs
fields that breaks the low energy
PQ symmetry (\ref{4*}) explicitly and softly. The 
corresponding mass parameter, $m_{12} = \sqrt{\lambda c_1} v_m$
should be of order 100 GeV to avoid fine tuning in the ordinary electroweak
Higgs sector. The second term in eq.(\ref{8*}), being inserted into 
eq.(\ref{5+}), induces mixing between axion and axial Higgs boson $A^0$
which is of order $m_{12}^2 v/v_m$. Hence, one expects the mixing angle
\[
   |\theta_{a,A^0}| \sim \frac{v}{v_m} 
          \cdot \frac{m_{12}^2}{|M_{A^0}^2 - M_a^2|}
\]
With $v_m \sim 10^4$ -- $10^5$ GeV 
this angle is in the range $10^{-2}$ -- $10^{-4}$,
but this estimate is again strongly parameter-dependent, and the 
mixing may be somewhat higher.

Thus, in our toy model the axion is a relatively light remnant of
extra strong interactions operating at very short distances. 
The axion mass may be
well within accessible range of energies; its interactions with 
ordinary matter come from mixing with axial Higgs boson 
$A^0$, and the mixing angle may not be negligibly small. The scalar 
potential of the ordinary Higgs fields exhibits softly broken PQ symmetry. 
We expect that these features are generic to the class of grand 
unified theories where the strong CP problem is solved in a way discussed in
this paper.

The author is indebted to G. Farrar for helpful and encouraging 
correspondence and to A. Dolgov, I. Khriplovich, V. Kuzmin, M.Shaposhnikov,
P. Tinyakov and M. Voloshin for useful discussions. This work was
supported in part by the Russian Foundation for Basic Research Grant No. 
96-02-17449a and
by the U.S. Civilian Research and Development Foundation for Independent 
States of FSU (CRDF) Award No. RP1-187.



\begin{thebibliography}{99}

\bibitem{Hooft} G. 't Hooft, {\em Phys. Rev. Lett.} {\bf 37} (1976) 172.
\bibitem{Reviews} N.V. Krasnikov, V.A. Matveev and A.N. Tavkhelidze,
{\em Elem. Chast. At. Yad.} {\bf 12} (1980) 100;
 J.E. Kim, {\em Phys. Rep.} {\bf 150} (1987) 1.
\bibitem{PQ} R.D. Peccei and H. Quinn, {\em Phys. Rev. Lett.} 
{\bf 38 } (1977)  1440.
\bibitem{WW} S. Weinberg, {\em Phys. Rev. Lett.} {\bf 40} (1978) 223;
             F. Wilczek, {\em Phys. Rev. Lett.} {\bf 40} (1978) 279.
\bibitem{invisible} J.E. Kim, {\em Phys. Rev. Lett.} {\bf 43} (1979) 103;
M. Shifman, A. Vainshtein and V. Zakharov,  {\em Nucl. Phys.} {\bf B166} 
(1980) 493;
A.R. Zhitnitsky, {\em Sov. J. Nucl. Phys.} {\bf 31} (1980) 260;
M. Dine, W. Fischler and M. Srednicki, {\em Phys. Lett.} {\bf B104} (1981) 199.
\bibitem{gravity} 
S. - J. Rey, {\em Phys. Rev.} {\bf D39} (1989) 3185;
R. Holman, S. Hsu, T. Kephart, E. Kolb, R. Watkins
and L. Widrow, {\em Phys. Lett.} {\bf B282} (1992) 132;
M. Kamionkowski and J. March-Russell, {\em Phys. Lett.} {\bf B282} (1992) 137;
S. Barr and D. Seckel, {\em Phys. Rev.} {\bf D46} (1992) 539;
S.M. Lusignioli and M. Roncadelli, {\em Phys. Lett.} {\bf B283} (1992) 278;
R. Kallosh, A. Linde, D. Linde and L. Susskind, {\em Phys. Rev.} {\bf D52} 
(1995) 912.
\bibitem{Tye} S.-H.H. Tye, {\em Phys. Rev. Lett.} {\bf 47} (1981) 1035.
\bibitem{Krasnikov} N.V. Krasnikov and V.A. Matveev, {\em Pisma Zh.E.T.F.} 
{\bf 35} (1982) 270.
\bibitem{Peskin} B. Holdom and M. Peskin, {\em Nucl. Phys.} {\bf B208} 
(1982) 397;
B. Holdom, {\em Phys. Lett.} {\bf B154} (1985) 316.
\bibitem{corrections} J. Ellis and M.K. Gaillard, {\em Nucl. Phys.} 
                      {\bf B150} (1979) 141;
I.B. Khriplovich, {\em Phys. Lett.} {\bf B173} (1986) 193;
I.B. Khriplovich and A.I Vainshtein, {\em Nucl. Phys.} {\bf B414} (1994) 27.

\end{thebibliography}
\end{document}